\documentclass[12pt]{article}
\pdfoutput=1
\usepackage{jheppub}
\usepackage{amsfonts}
\usepackage{amsmath,amssymb} % for mathematical symbols
\usepackage{graphicx} % for \includegraphics command and options
\usepackage[dvipsnames,x11names]{xcolor} % to use color

\usepackage[nottoc,notlof,notlot]{tocbibind} % to put the bibliography in the ToC
\usepackage[titles]{tocloft} % to alter the style of the ToC
\usepackage{hyperref}
\usepackage{cleveref}
\usepackage{float}
\usepackage{subcaption}
\numberwithin{equation}{section} % Equations numbered as <section>.#

\title{\textbf{Traversable wormhole without interaction}}

\author{Nilakash Sorokhaibam}\emailAdd{nilakash@niser.ac.in}

\affiliation{
National Institute of Science Education and Research, HBNI, Bhubaneswar 752050, Odisha, India}

  \abstract{We show that strong quantum entanglement can support a stable traversable wormhole without any explicit interaction or tunnelling term between the two boundary theories of the wormhole. Specifically we work with two complex SYK models. The entangled state is prepared using a tunnelling term in imaginary time but the tunnelling term is removed from the time evolution operator so the two complex SYK models are not coupled. Low temperature states show revival dynamics which is the hallmark of a traversable wormhole geometry. To send any meaningful information from one system to the other, one only needs to turn on a very small interaction term. The technique that we are employing can be applied to other systems to study aspects of quantum entanglement.}

\begin{document}
\maketitle

%%%%%%%%%%%%%%%%%%%%%%%%%%%%%%%%%%%%%%%%%% 
\section{Introduction and Summary}
\label{intro}
Recent works have shown that quantum effects can stabilize traversable wormholes \cite{Gao:2016bin,Maldacena:2017axo,Maldacena:2018gjk}. These developments has also been extended to boundary theories. The wormhole physics in the boundary theories has been most concretely realized using Sachdev-Ye-Kiteav(SYK) models \cite{Sachdev:1993,Kitaev:2015}. In recent years, SYK models have been studied extensively due to its maximally chaotic nature and the belief that it is dual to some quantum gravity theory.

SYK models are theories with a large number of fermions which are all-to-all coupled and the couplings are random variables drawn from a Gaussian ensemble. With respect to wormhole physics, two copies of SYK models are considered \cite{Maldacena:2018lmt, Garcia-Garcia:2019poj, Maldacena:2019ufo, Plugge:2020wgc, Qi:2020ian, sahoo2020traversable}. The two systems(say L and R, for left and right) are coupled using an interaction or  tunnelling term. It has been shown that the full system undergoes a first order phase transition from a chaotic phase to a Fermi liquid phase. The chaotic phase is dual to two blackhole phase and the Fermi liquid phase is dual to a traversable wormhole. From real time dynamics, the most remarkable feature of the wormhole phase is that the insertion of a particle in one side or a SYK system is scrambled but within a characteristic time (depending on the strength of the tunnelling term) the particle is unscrambled on the other side or the other SYK system and the process goes on and on.

The revival dynamics is captured by the two point functions $\langle \psi_L(t) \psi^{\dagger}_R\rangle$ and $\langle \psi_L(t) \psi^{\dagger}_L\rangle$. Consider the state of the wormhole to be $|\Psi_{WH}\rangle$. The first two-point function measures the overlap between the state $\psi^{\dagger}_R|\Psi_{WH}\rangle$ and the state $\psi^{\dagger}_L(t)|\Psi_{WH}\rangle$ after a time $t$. In the traversable wormhole phase, the overlap oscillates. When $\langle \psi_L(t) \psi^{\dagger}_R\rangle$ is large, $\langle \psi_L(t) \psi^{\dagger}_L\rangle$ is small and vice versa. In all the works so far, the interaction or tunnelling term is absolutely necessary to realize the wormhole geometry.

In this work, we will show that strong quantum entanglement alone can support a stable traversable wormhole without any explicit interaction between the two boundary theories. But note that the entanglement between the two sides should be strong enough to take the total system to the wormhole phase, otherwise the total system will be in the two blackhole phase. Also note that to send any meaningful information from one side to the other side, one needs to turn on a small interaction or tunnelling term. This is because of the well known fact that entanglement alone cannot be used to transmit any meaningful information. But once the traversable wormhole is set up using entanglement alone, the interaction strength required for information transmission is very small.

We will consider two complex SYK models and couple them using a tunnelling term.
\begin{gather}
\label{H2S}
H_{2S}=H_{SYK,L}+H_{SYK,R}\\
\label{Htot}
H_{tot}=H_{SYK,L}+H_{SYK,R}+i\mu H_{int}\\
H_{SYK,L(R)}=\sum_{i,j,k,l}j_{ij,kl}\psi^\dagger_{iL(R)}\psi^\dagger_{jL(R)}\psi_{kL(R)}\psi_{lL(R)}, \qquad H_{int}=\left(\psi^\dagger_L\psi_R+\psi_L\psi^\dagger_R\right)\
\end{gather}
The Hamiltonian $H_{2S}$ is simply the sum of two complex SYK Hamiltonians. $H_{tot}$ is the Hamiltonian with the tunnelling term considered in \cite{sahoo2020traversable}. Note that we have explicitly written $i\mu$ in $H_{tot}$, not as a factor in $H_{int}$. $j_{ij,kl}$ in both $H_{SYK,L}$ and $H_{SYK,R}$ are same. $j_{ij,kl}$ are the well-known disordered couplings drawn from a Gaussian ensemble. $H_{int}$ commutes with both $H_{SYK,L}$ and $H_{SYK,R}$. So, $\langle H_{int}\rangle=Q_{int}$ is a conserved quantity.

%The tunnelling term in \cite{Maldacena:2018lmt} does not commute with the Hamiltonian of the original SYK model with Majorana fermions.

Since we are dealing with complex fermions, the fermion commutation relations force
\begin{eqnarray}
j_{ij,kl}=-j_{ji,kl}=-j_{ij,lk}=j_{kl,ij}\
\end{eqnarray}
The Hamiltonian $H_{tot}$ also has a mirror symmetry. Simultaneously interchanging
\begin{eqnarray}
\psi_{iL}\to\psi_{iR}, \quad \psi^\dagger_{iL}\to\psi^\dagger_{iR}, \quad \psi^\dagger_{iR}\to-\psi^\dagger_{iL}, \quad \psi_{iR}\to-\psi_{iL}\
\label{msym}
\end{eqnarray}
leave the Hamiltonian unchanged.

In the study of thermal quantum field theories using imaginary time(path integral) formalism, a conserved charge is a part of the Hamiltonian. If one perform a Wick rotation to obtain real time quantities, the time evolution operator will include the charge term as a part of the Hamiltonian. But in case of real time dynamics, one can explicitly turn on the charge term as part of the thermal state but the time evolution does not have the charge term as part of the Hamiltonian. The simplest example is the difference between mass and charge of a single free fermion considered in Appendix B of \cite{sorokhaibam2019phase}.

In the same spirit, our aim is to calculate the two point functions of the uncoupled (two SYK) system with Hamiltonian $H_{2S}$ but the thermal state have non-zero $Q_{int}$ charge. To be precise, the thermal state we will be considering has the density matrix
\begin{equation}
\rho_{in}=e^{-\beta H_{tot}}=e^{-\beta(H_{SYK,L}+H_{SYK,R}+i\mu H_{int})}\
\label{inden}
\end{equation}
The $L$ and $R$ systems are entangled. The strength of the entanglement depends on the parameter $\mu$. Larger $\mu$ means stronger entanglement. It has been shown that this state is close to the thermofield double state at low temperature \cite{Maldacena:2018lmt,sahoo2020traversable}. It has been shown that the system with the Hamiltonian $H_{tot}$ undergoes a first order phase transition. It is like Hawking-Page transition. The high temperature chaotic phase is dual to two black holes while the low temperature Fermi liquid phase is dual to an eternally traversable wormhole. In the low temperature wormhole phase, the system exhibits revival dynamics.

Our main result is that we find revival dynamics when we time evolve using the Hamiltonian $H_{2S}$ which does not have the tunnelling term but starting from the initial state with the density matrix given by (\ref{inden}). The $e^{-\beta(i\mu H_{int})}$ factor in the density matrix entangles the two SYK systems. But the time evolution operator is
\begin{equation}
U_{2S}(t)=e^{-itH_{2S}}\
\label{U2S}
\end{equation}

If we instead use the evolution operator
\begin{equation}
\label{Utot}
U_{tot}(t)=e^{-itH_{tot}}\
\end{equation}
then the two SYK systems would be explicitly coupled with the tunnelling term $i\mu H_{int}$. 

Our convention of the two point functions are
\begin{gather}
G^<_{ab}(t_1-t_2)=G^<_{ab}(t_1,t_2)=i\langle\psi^{\dagger}_b(t_2)\psi_a(t_1)\rangle\\
G^>_{ab}(t_1-t_2)=G^>_{ab}(t_1,t_2)=-i\langle\psi_a(t_1)\psi^{\dagger}_b(t_2)\rangle\\
G^R_{ab}(t_1-t_2)=G^R_{ab}(t_1,t_2)=\Theta(t_1-t_2)\left(G^>_{ab}(t_1,t_2)-G^<_{ba}(t_1,t_2)\right)\
\end{gather}
where $a,b=L,R$. The revival dynamics will be examined using the transmission amplitude $T_{LR}(t)$ and the return amplitude $T_{LL}(t)$ defined by
\begin{equation}
T_{LR}(t)=2*|G^>_{LR}(t)|, \qquad T_{LL}(t)=2*|G^>_{LL}(t)|\
\end{equation}

We will differentiate the fermion evolutions using (\ref{U2S}) and (\ref{Utot}) as follows
\begin{gather}
\label{tpsi_evo}
\tilde{\psi}_a(t)=U_{2S}^{\dagger}(t)\psi_a(0)U_{2S}(t), \quad \tilde{\psi}_a^{\dagger}(t)=U_{2S}^{\dagger}(t)\psi^{\dagger}_a(0)U_{2S}(t)\\
\label{psi_evo}
\psi_a(t)=U_{tot}^{\dagger}(t)\psi_a(0)U_{tot}(t), \quad \psi^{\dagger}_a(t)=U_{tot}^{\dagger}(t)\psi^{\dagger}_a(0)U_{tot}(t)\
\end{gather}
Note that $\tilde{\psi}_a(0)=\psi_a(0)$. The commutation relation between $H_{int}$ and the microscopic fermionic operators are
\begin{gather}
\label{commut1}
\left[H_{int},\psi_L\right]=\psi_R, \quad \left[H_{int},\psi_L\right]=\psi_R\\
\label{commut2}
\left[H_{int},\psi^{\dagger}_L\right]=-\psi^{\dagger}_R, \quad \left[H_{int},\psi^{\dagger}_R\right]=-\psi^{\dagger}_L\
\end{gather}
Using these commutation relations, we obtain the following BCH-like relations.
\begin{eqnarray}
\label{psitrans}
\psi_{L(R)}(t)=e^{-t\mu H_{2S}}\tilde{\psi}_{L(R)}(t)e^{t\mu H_{2S}}=\frac{1}{\alpha}\left[\cosh(\mu t)\tilde{\psi}_{L(R)}(t)-\sinh(\mu t)\tilde{\psi}_{R(L)}(t)\right]\\
\label{psidtrans}
\psi^{\dagger}_{L(R)}(t)=e^{-t\mu H_{2S}}\tilde{\psi}^{\dagger}_{L(R)}(t)e^{t\mu H_{2S}}=\frac{1}{\alpha}\left[\cosh(\mu t)\tilde{\psi}^{\dagger}_{L(R)}(t)+\sinh(\mu t)\tilde{\psi}^{\dagger}_{R(L)}(t)\right]\\
\label{tpsitrans}
\tilde{\psi}_{L(R)}(t)=e^{t\mu H_{2S}}\psi_{L(R)}(t)e^{-t\mu H_{2S}}=\frac{1}{\alpha}\left[\cosh(\mu t)\psi_{L(R)}(t)+\sinh(\mu t)\psi_{R(L)}(t)\right]\\
\label{tpsidtrans}
\tilde{\psi}^{\dagger}_{L(R)}(t)=e^{t\mu H_{2S}}\psi^{\dagger}_{L(R)}(t)e^{-t\mu H_{2S}}=\frac{1}{\alpha}\left[\cosh(\mu t)\psi^{\dagger}_{L(R)}(t)-\sinh(\mu t)\psi^{\dagger}_{R(L)}(t)\right]\
\end{eqnarray}
We have used the fact that $H_{int}$ commutes with $H_{2S}$, so $e^{-it H_{tot}}=e^{-it H_{2S}+\mu tH_{int}}=e^{-it H_{2S}}e^{\mu tH_{int}}=e^{\mu tH_{int}}e^{-it H_{2S}}$. The normalization constant $\alpha$ is fixed using the relation
\begin{eqnarray}
\psi_L(t)\psi^{\dagger}_R(t)&=&\frac{\cosh(\mu t)^2+\sinh(\mu t)^2}{\alpha^2}\,\tilde{\psi}_L(t)\tilde{\psi}^{\dagger}_R(t)\nonumber\\
\Rightarrow \qquad \qquad \alpha &=& \sqrt{\cosh(\mu t)^2+\sinh(\mu t)^2}\
\label{alpha}
\end{eqnarray}
Now using (\ref{tpsitrans},\ref{alpha}), the two point functions of the uncoupled system in terms of two-point functions of the coupled system in the same state are given by
\begin{eqnarray}
\label{tGGLL}
\tilde{G}^>_{LL}(t)=\frac{1}{\sqrt{\cosh(\mu t)^2+\sinh(\mu t)^2}}\,\left(\cosh(\mu t)G^>_{LL}(t)+\sinh(\mu t) G^>_{LR}(t)\right)\\
\label{tGGLR}
\tilde{G}^>_{LR}(t)=\frac{1}{\sqrt{\cosh(\mu t)^2+\sinh(\mu t)^2}}\,\left(\cosh(\mu t)G^>_{LR}(t)+\sinh(\mu t) G^>_{LL}(t)\right)\
\end{eqnarray}
where for the last relation we have used the mirror symmetry (\ref{msym}) which implies that $G^>_{LL}(t_1,t_2)=G^>_{RR}(t_1,t_2)$. We will numerically solve for the solutions of $G^>_{LL}(t)$ and $G^>_{LR}(t)$ in the state with the density matrix given by (\ref{inden}). The transmission amplitude $\tilde{T}_{LR}(t)$ and the return amplitude $\tilde{T}_{LL}(t)$ for the uncoupled system are defined by
\begin{equation}
\tilde{T}_{LR}(t)=2*|\tilde{G}^>_{LR}(t)|, \qquad \tilde{T}_{LL}(t)=2*|\tilde{G}^>_{LL}(t)|\
\end{equation}

The mirror symmetry and the fact that the thermal state is uncharged (with U(1) charge,  $G^>_{LL}(0)=-G^<_{LL}(0)=-\frac{i}{2}$) imply that the following relations hold for the two-point functions.
\begin{gather}
\label{Gsym1}
G^{>(<)}_{LL}(t)=G^{>(<)}_{RR}(t), \quad G^>_{LL}(t)=-G^>_{LL}(-t)^*\\
\label{Gsym2}
G^<_{LL}(t)=G^>_{LL}(t)^*=-G^>_{LL}(-t)\\
\label{Gsym3}
G^{>(<)}_{LR}(t)=-G^{>(<)}_{RL}(t), \quad G^>_{LR}(t)=G^>_{LR}(-t)^*\\
\label{Gsym4}
G^<_{LR}(t)=G^>_{LR}(t)^*=G^>_{LR}(-t)\
\end{gather}
These relations will significantly simplify the numerical task at hand. We only have to consider $G^>_{LL}(t)$ and $G^>_{LR}(t)$ for $t\geq 0$.

The reason why we do not consider the Majorana SYK model of \cite{Maldacena:2018lmt} is because the tunnelling term does not commute with the SYK Hamiltonian. So it would be an intractable exercise to remove the tunnelling term from the full time evolution operator.

The technical procedures we will be following are very simple. We will first consider two coupled complex SYK models with the tunnelling term  and calculate the two-point functions. We will then systematically remove the tunnelling term from the time evolution operator of the two-point functions using the simple relations (\ref{tGGLL},\ref{tGGLR}). This will give the two-point functions of the system with the two sides decoupled. But the two sides are entangled.

The numerical parts of this work was performed in a laptop computer with Intel Core i3-7020U processor, taking around 10 hours of time for all the different parameter ranges considered. The numerical tasks were not memory intensive, requiring around 20 Megabytes of memory.

\section{Coupled complex SYK models}
In this section we will consider the system with the coupled Hamiltonian \ref{Htot}. We will derive the Schwinger-Dyson(SD) equations in real time from which we can numerically calculate the real time two-point functions. The action in the Keldysh contour is
\begin{eqnarray}
S=\int_{\mathcal{C}} dt \left(i\psi^\dagger_L\partial_t \psi_L+i\psi^\dagger_R\partial_t \psi_R-H_{tot}\right)\
\end{eqnarray}
We can write the SYK couplings $j_{ij,kl}$ into real and imaginary parts.
\begin{eqnarray}
&& j_{ij,kl}\psi^\dagger_{iL(R)}\psi^\dagger_{jL(R)}\psi_{kL(R)}\psi_{lL(R)}+j^*_{kl,ij}\psi^\dagger_{kL(R)}\psi^\dagger_{lL(R)}\psi_{iL(R)}\psi_{jL(R)}\nonumber\\
&=& j_{Re;ij,kl}\left(\psi^{\dagger}_{iL(R)} \psi^{\dagger}_{jL(R)} \psi_{kL(R)} \psi_{lL(R)} + \psi^{\dagger}_{kL(R)} \psi^{\dagger}_{lL(R)} \psi_{iL(R)} \psi_{jL(R)}\right)\nonumber\\
&&\qquad +i\,j_{Im;ij,kl}\,\left(\psi^{\dagger}_{iL(R)} \psi^{\dagger}_{jL(R)} \psi_k \psi_{lL(R)} - \psi^{\dagger}_{kL(R)} \psi^{\dagger}_{lL(R)} \psi_{iL(R)} \psi_{jL(R)}\right)
\end{eqnarray}
where $j_{Re;ij,kl}$ and $j_{Im;ij,kl}$ are the real and imaginary parts of $j_{ij,kl}$. $j_{Re;ij,kl}$ and $j_{Re;ij,kl}$ are real numbers drawn from a Gaussian ensemble with variance $J^2$. After performing the disorder averaging, the partition function is
\begin{eqnarray}
Z&=&\int \mathcal{D}\psi^{\dagger}_L\mathcal{D}\psi^{\dagger}_R\mathcal{D}\psi_L\mathcal{D}\psi_R \exp\left[\int_{\mathcal{C}} dt \, \sum_i \begin{pmatrix}
\psi^\dagger_{iL} & \psi^\dagger_{iR} \end{pmatrix}\,\begin{pmatrix}
-\partial_t & -\mu\\
\mu & -\partial_t \end{pmatrix}\,\begin{pmatrix}
\psi_{iL} \\ \psi_{iR} \end{pmatrix}\right.\nonumber\\
&& \left. + \int dt_1 dt_2 \, \frac{J^2}{N^3}\sum_{i,j,k,l}\left(P_{+,ijkl}(t_1)P_{+,ijkl}(t_2)+P_{-,ijkl}(t_1)P_{-,ijkl}(t_2)\right)\right]\\
P_{\pm,ijkl}&=&\psi^\dagger_{iL}\psi^\dagger_{jL}\psi_{kL}\psi_{lL}\pm\psi^\dagger_{kL}\psi^\dagger_{lL}\psi_{iL}\psi_{jL}+\psi^\dagger_{iR}\psi^\dagger_{jR}\psi_{kR}\psi_{lR}\pm\psi^\dagger_{kR}\psi^\dagger_{lR}\psi_{iR}\psi_{jR}\nonumber\
\end{eqnarray}
The averaged contour ordered propagators are defined as
\begin{eqnarray}
G^F_{L(R)L(R)}(t_1,t_2)=-\frac{i}{N}\sum_i\langle\mathcal{T}_{\mathcal{C}}\,\left(\psi_{iL(R)}(t_1)\psi^\dagger_{iL(R)}(t_2)\right)\rangle\
\end{eqnarray}
We will enforce these relations using Lagrange multipliers $\Sigma$'s. So the partition function becomes
\begin{eqnarray}
Z&=&\int \prod_{a=L,R}\mathcal{D}\psi^{\dagger}_a\mathcal{D}\psi_a \prod_{a,b=L,R}\mathcal{D}\Sigma_{ab} \, \exp\left[\int_{\mathcal{C}} dt \, \sum_i\begin{pmatrix}
\psi^\dagger_{iL} & \psi^\dagger_{iR} \end{pmatrix}\,\begin{pmatrix}
-\partial_t-i\Sigma_{LL} & -\mu-i\Sigma_{LR}\\
\mu-i\Sigma_{RL} & -\partial_t-i\Sigma_{RR} \end{pmatrix}\,\begin{pmatrix}
\psi_{iL} \\ \psi_{iR} \end{pmatrix}\right.\nonumber\\
&& \left. \qquad\qquad + \int_{\mathcal{C}} dt_1 dt_2 \, \sum_{a,b=L,R}\left(N\Sigma_{ab}(t_1,t_2)G_{ba}(t_2,t_1)-\frac{J^2 N}{4}G_{ab}(t_1,t_2)^2G_{ba}(t_2,t_1)^2\right)\right]\
\end{eqnarray}
Integrating out the quadratic terms of the fermions, we get the effective action
\begin{eqnarray}
\frac{iS_{eff}}{N}&=& \log \det \begin{pmatrix}
\partial_t+i\Sigma_{LL} & \mu+i\Sigma_{LR}\\
-\mu+i\Sigma_{RL} & \partial_t+i\Sigma_{RR} \end{pmatrix}\nonumber\\
&& + \int dt_1 dt_2 \, \sum_{a,b=L,R}\left(\Sigma_{ab}(t_1,t_2)G_{ba}(t_2,t_1)-\frac{J^2}{4}G_{ab}(t_1,t_2)^2G_{ba}(t_2,t_1)^2\right)\nonumber\\
\end{eqnarray}
The equations of motion of $G_{LL}$, $G_{LR}$, $\Sigma_{LL}$ and $\Sigma_{LR}$ are
\begin{gather}
\label{cSD1}
G^R_{LL}(\omega)=\frac{\left(\omega-\Sigma^R_{LL}(\omega)\right)}{(\omega-\Sigma^R_{LL}(\omega))^2-(\mu+i\Sigma^R_{LR}(\omega))^2}\\
\label{cSD2}
G^R_{LR}(\omega)=\frac{-i\mu+\Sigma^R_{LR}(\omega)}{(\omega-\Sigma^R_{LL}(\omega))^2-(\mu+i\Sigma^R_{LR}(\omega))^2}\\
\label{cSD3}
\Sigma^{>(<)}_{LL}(t_1,t_2)=J^2G^{>(<)}_{LL}(t_1,t_2)^2G^{<(>)}_{LL}(t_2,t_1)\\
\label{cSD4}
\Sigma^{>(<)}_{LR}(t_1,t_2)=-J^2G^{>(<)}_{LR}(t_1,t_2)^2G^{<(>)}_{LR}(t_2,t_1)\
\end{gather}
where we have used the relations (\ref{Gsym1},\ref{Gsym3}). These are the Schwinger-Dyson equations for the two coupled complex SYK models. In the next section we will solve these equations numerically. When $\mu=0$ and the two SYK systems are uncoupled, $G_{LR}(t_1,t_2)=0$ is one of the consistent solutions and the only physical solution when the two systems are not entangled. On the other hand, it is not possible to solve the entangled but uncoupled system directly using the SD equations, the numerics will always converge to $G_{LR}(t_1,t_2)=0$. This is where our operator algebra technique elaborated in section {\ref{intro} comes in handy.

\section{Revival dynamics in uncoupled SYK models}
In this section, we will show the revival dynamics in the uncoupled system purely due to quantum entanglement. But first we will have to solve the SD equations of the two coupled complex SYK models. We will set $J=1$. We use weighted iteration method \cite{Maldacena:2016hyu} to solve (\ref{cSD1}), (\ref{cSD2}), (\ref{cSD3}) and (\ref{cSD4}). For the numerical implementation we also need the relations
\begin{gather}
\Sigma^R_{ab}(t_1,t_2)=\Theta(t_1-t_2)\left[\Sigma^>_{ab}(t_1,t_2)-\Sigma^<_{ab}(t_1,t_2)\right]\\
G^>_{ab}(\omega)=-\,\frac{i}{1+e^{-\beta\omega}}\,A_{ab}(\omega)\\
G^<_{ab}(\omega)=\frac{i}{1+e^{\beta\omega}}\,A_{ab}(\omega)\\
A_{LL}(\omega)=-2\;\text{Im}\,G^R_{LL}(\omega)\\
A_{LR}(\omega)=2i\;\text{Re}\,G^R_{LR}(\omega)\
\end{gather}
where $A_{ab}(\omega)$'s are the spectral functions. To initiate the first iteration, we use the real time solution of the solvable $(q=2)$ SYK model which is given by
\begin{equation}
A(\omega)=\frac{1}{J_2}\,\sqrt{4J^2_2-\omega^2}, \qquad \omega \in \{-2J^2_2,2J_2^2\}\
\end{equation}
This is done at high temperature $\beta=10$. Once we obtain the solutions for our coupled systems, we used the spectral functions to further solve the SD equations at lower temperature. We check for convergence as in \cite{sorokhaibam2019phase} by calculating
\begin{equation}
\Delta A_{LL}=\sum_{\omega}|A_{LL}(\omega)-A^{prev}_{LL}(\omega)|\
\end{equation}
Usually this quantity decreases monotonically. When this quantity decreases below a preset tolerance limit, we declare that we have obtained the solutions. But note that as in \cite{sorokhaibam2019phase}, when the system undergoes the phase transition, $\Delta A_{LL}$ jumps a bump. It first decreases to some extend then it will increase for some iterations and finally it will rapidly converge to the solutions of the wormhole phase.

We used the frequency range $\{-10^5*\Delta\omega,10^5*\Delta\omega\}$ where $\Delta\omega=5\times 10^{-5}$. We used the real time range $\{-5000\times\Delta t, 5000\times \Delta t\}$ where $\Delta t=1$. Using the relations (\ref{Gsym1},\ref{Gsym2},\ref{Gsym3},\ref{Gsym4}), we only have to calculate $G^>_{LL}(t)$ and $G^>_{LR}(t)$ for $t\geq 0$. Moreover we also have
\begin{gather}
\label{ALLsym}
A_{LL}(-\omega)=A_{LL}(\omega)\\
A_{LR}(-\omega)=-A_{LR}(\omega)\
\end{gather}
So further we only need to compute $A_{LL}(\omega)$ and $A_{LR}(\omega)$ for $\omega\geq 0$. For $\mu=0.05$, the phase transition occurs while the system is cooling down from $\beta=70$ to $\beta=80$. This agrees with a rough estimate from Figure 3(c) of \cite{sahoo2020traversable}. For $\mu=0.25$, the phase transition occurs when the system is cooled down from $\beta=200$ to $\beta=300$. For $\mu=0.05$, $G^>_{LL}(t)$ and $G^>_{LR}(t)$(so also $T_{LR}(t)$ and $T_{LL}(t)$) do not change significantly when we change the temperature from $\beta=100$ to $\beta=200$ in the wormhole phase.

We will present results for $\mu=0.05, \beta=100$ and $\mu=0.025, \beta=300$. Figure \ref{fig:GG_mu005_b100} is the plots of $G^>_{LL}(t)$ and $G^>_{LR}(t)$ of the coupled system with the Hamiltonian $H_{tot}$ (\ref{Htot}) with $\mu=0.025, \beta=300$ in the wormhole phase. Figure \ref{fig:spectralfunc} is the plots of the spectral functions $A_{LL}(\omega)$ and $A_{LR}(\omega)$ for two sets of parameters. Figure \ref{fig:amps} is the plots of the transmission amplitude $T_{LR}(t)$ and return amplitude $T_{LL}(t)$. For comparison we have also plotted $T_{SYK}=2*|G^>(t)|$ for the complex SYK model which is always in the chaotic/blackhole phase.
\begin{figure}[h]
\centering
\begin{subfigure}{.5\textwidth}
  \centering
  \includegraphics[width=.9\linewidth]{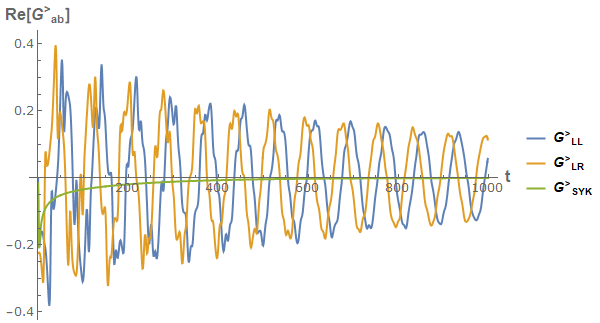}
  \caption{Real parts of $G^>(t)$'s.}
  \label{fig:ReGG_mu0025_b300}
\end{subfigure}%
\begin{subfigure}{.5\textwidth}
  \centering
  \includegraphics[width=.9\linewidth]{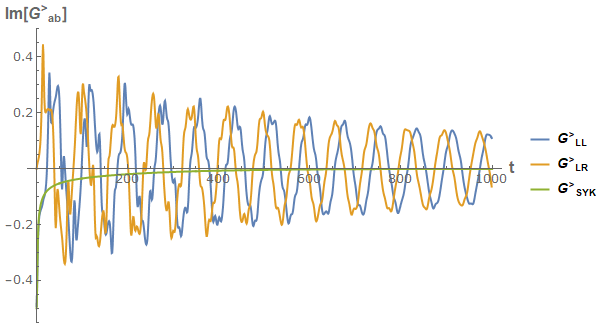}
  \caption{Imaginary parts of $G^>(t)$'s.}
  \label{fig:ImGG_mu0025_b300}
\end{subfigure}
\caption{\small{The real and imaginary parts of $G^>_{LL}(t)$ and $G^>_{LR}(t)$ for $\mu=0.025, \beta=300$.}}
\label{fig:GG_mu0025_b300}
\end{figure}

\begin{figure}[h]
\centering
\begin{subfigure}{.5\textwidth}
  \centering
  \includegraphics[width=.9\linewidth]{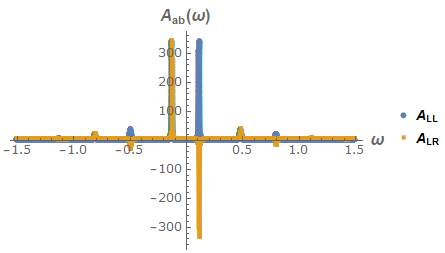}
  \caption{$\mu=0.05, \beta=100$}
  \label{fig:spectralfunc_mu005_b100}
\end{subfigure}%
\begin{subfigure}{.5\textwidth}
  \centering
  \includegraphics[width=.9\linewidth]{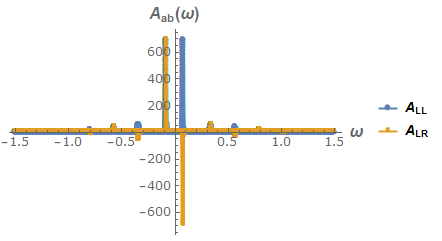}
  \caption{$\mu=0.025, \beta=300$}
  \label{fig:spectralfunc_mu0025_b300_2}
\end{subfigure}
\caption{\small{Spectral functions.}}
\label{fig:spectralfunc}
\end{figure}

\begin{figure}[h]
\centering
\begin{subfigure}{.5\textwidth}
  \centering
  \includegraphics[width=.9\linewidth]{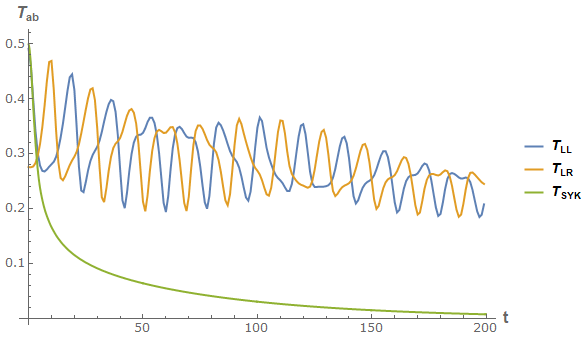}
  \caption{$\mu=0.05$}
  \label{fig:amps_mu005_b100}
\end{subfigure}%
\begin{subfigure}{.5\textwidth}
  \centering
  \includegraphics[width=.9\linewidth]{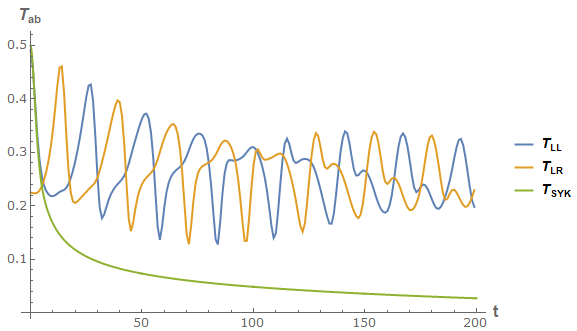}
  \caption{$\mu=0.025$}
  \label{fig:amps_mu005_b100}
\end{subfigure}
\caption{\small{Transmission and return amplitudes.}}
\label{fig:amps}
\end{figure}

Now for the uncoupled system with the Hamiltonian $H_{2S}$ (\ref{H2S}) in the entangled state with density matrix \ref{inden}, we used the relations (\ref{tGGLL},\ref{tGGLR}) to calculate the two-point functions. Figure \ref{fig:GG_mu0025_b300} is the plots of the $G^>_{LL}(t)$ and $G^>_{LR}(t)$  for the two sets of parameters.
\begin{figure}[h]
\centering
\begin{subfigure}{.5\textwidth}
  \centering
  \includegraphics[width=.9\linewidth]{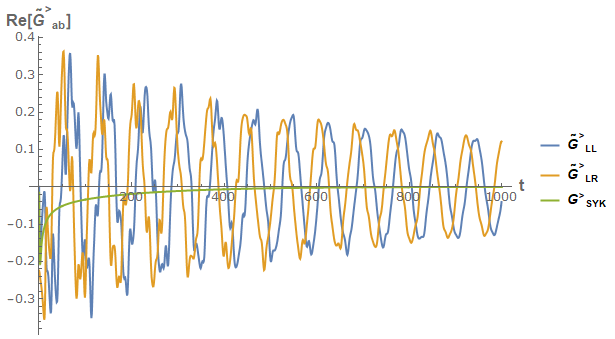}
  \caption{Real parts of $\tilde{G}^>(t)$'s.}
  \label{fig:ReGG_mu0025_b300}
\end{subfigure}%
\begin{subfigure}{.5\textwidth}
  \centering
  \includegraphics[width=.9\linewidth]{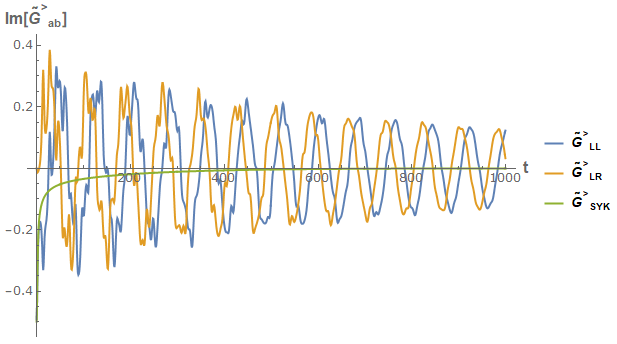}
  \caption{Imaginary parts of $\tilde{G}^>(t)$'s.}
  \label{fig:ImGG_mu0025_b300}
\end{subfigure}
\caption{\small{The real and imaginary parts of $\tilde{G}^>_{LL}(t)$ and $\tilde{G}^>_{LR}(t)$ for $\mu=0.025, \beta=300$.}}
\label{fig:GG_mu0025_b300}
\end{figure}
Our main result is shown in Figure \ref{fig:tamps_mu005_b100}. It is the plots of the transmission amplitude $T_{LR}(t)$ and return amplitude $T_{LL}(t)$ in the uncoupled system in the state with the density matrix (\ref{inden}).
\begin{figure}[h]
\centering
\begin{subfigure}{.5\textwidth}
  \centering
  \includegraphics[width=.9\linewidth]{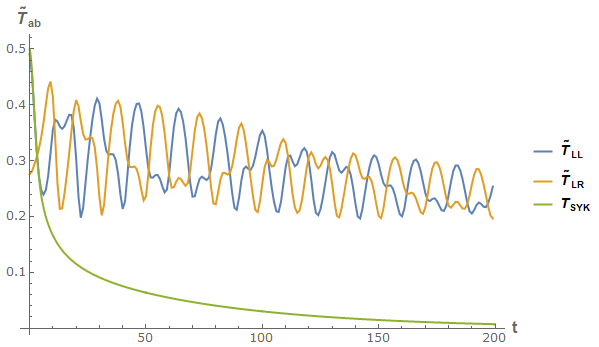}
  \caption{$\mu=0.05, \beta=100$}
  \label{fig:tamps_mu005_b100}
\end{subfigure}%
\begin{subfigure}{.5\textwidth}
  \centering
  \includegraphics[width=.9\linewidth]{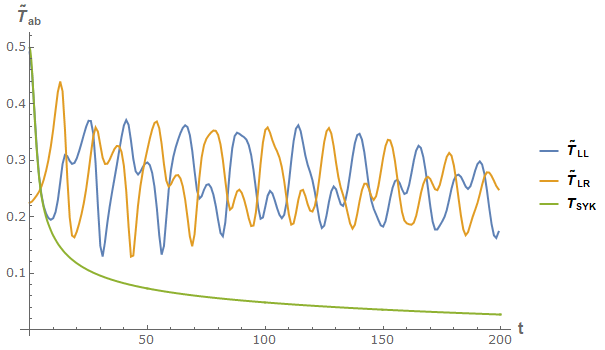}
  \caption{$\mu=0.025, \beta=300$}
  \label{fig:tamps_mu0025_b300}
\end{subfigure}
\caption{\small{The transmission amplitude $\tilde{T}_{LL}(t)$ and the return amplitude $\tilde{T}_{LR}(t)$ in the uncoupled system and $\tilde{T}_{LL}(t)$ in the chaotic SYK model.}}
\label{fig:tamps}
\end{figure}
\section{Conclusions and Discussions}
\label{cnd}
We show that strong quantum entanglement can support a stable traversable wormhole without any explicit interaction or tunnelling term between the two boundary theories of the wormhole. Specifically we work with two complex SYK models. The entangled state is prepared using an interaction term in imaginary time but the interaction term is removed from the time evolution operator so the two complex SYK models are not coupled. Low temperature states show revival dynamics which is the hallmark of traversable wormhole geometry. The entanglement has to be strong enough to take the system to the traversable wormhole phase.  To send any meaningful information from one system to the other, one only needs to turn on a very small interaction term. The technique that we are employing can be applied to other systems to study aspects of quantum entanglement.

\section*{Acknowledgement}
This work was started while the author was in quarantine due to COVID-19 pandemic at Victory High School, Kakching, Manipur (India). The author thanks the management team of the community quarantine centre for the comfortable stay.
\bibliography{sykquench} 

\providecommand{\href}[2]{#2}\begingroup\raggedright\begin{thebibliography}{10}

\bibitem{Gao:2016bin}
P.~Gao, D.~L. Jafferis, and A.~C. Wall, {\it {Traversable Wormholes via a
  Double Trace Deformation}},  {\em JHEP} {\bf 12} (2017) 151,
  [\href{http://arxiv.org/abs/1608.05687}{{\tt arXiv:1608.05687}}].

\bibitem{Maldacena:2017axo}
J.~Maldacena, D.~Stanford, and Z.~Yang, {\it {Diving into traversable
  wormholes}},  {\em Fortsch. Phys.} {\bf 65} (2017), no.~5 1700034,
  [\href{http://arxiv.org/abs/1704.05333}{{\tt arXiv:1704.05333}}].

\bibitem{Maldacena:2018gjk}
J.~Maldacena, A.~Milekhin, and F.~Popov, {\it {Traversable wormholes in four
  dimensions}},  \href{http://arxiv.org/abs/1807.04726}{{\tt
  arXiv:1807.04726}}.

\bibitem{Sachdev:1993}
S.~Sachdev and J.~Ye, {\it Gapless spin-fluid ground state in a random quantum
  heisenberg magnet.},  {\em Phys. Rev. Lett.} {\bf 70} (1993) 3339.

\bibitem{Kitaev:2015}
A.~Y. Kitaev, {\it Entanglement in strongly-correlated quantum matter}, . Talk
  at KITP, University of California, Santa Barbara.

\bibitem{Maldacena:2018lmt}
J.~Maldacena and X.-L. Qi, {\it {Eternal traversable wormhole}},
  \href{http://arxiv.org/abs/1804.00491}{{\tt arXiv:1804.00491}}.

\bibitem{Garcia-Garcia:2019poj}
A.~M. García-García, T.~Nosaka, D.~Rosa, and J.~J. Verbaarschot, {\it
  {Quantum chaos transition in a two-site Sachdev-Ye-Kitaev model dual to an
  eternal traversable wormhole}},  {\em Phys. Rev. D} {\bf 100} (2019), no.~2
  026002, [\href{http://arxiv.org/abs/1901.06031}{{\tt arXiv:1901.06031}}].

\bibitem{Maldacena:2019ufo}
J.~Maldacena and A.~Milekhin, {\it {SYK wormhole formation in real time}},
  \href{http://arxiv.org/abs/1912.03276}{{\tt arXiv:1912.03276}}.

\bibitem{Plugge:2020wgc}
S.~Plugge, E.~Lantagne-Hurtubise, and M.~Franz, {\it {Revival dynamics in a
  traversable wormhole}},  {\em Phys. Rev. Lett.} {\bf 124} (2020), no.~22
  221601, [\href{http://arxiv.org/abs/2003.03914}{{\tt arXiv:2003.03914}}].

\bibitem{Qi:2020ian}
X.-L. Qi and P.~Zhang, {\it {The Coupled SYK model at Finite Temperature}},
  {\em JHEP} {\bf 05} (2020) 129, [\href{http://arxiv.org/abs/2003.03916}{{\tt
  arXiv:2003.03916}}].

\bibitem{sahoo2020traversable}
S.~Sahoo, Étienne Lantagne-Hurtubise, S.~Plugge, and M.~Franz, {\it
  Traversable wormhole and hawking-page transition in coupled complex syk
  models},  2020.

\bibitem{sorokhaibam2019phase}
N.~Sorokhaibam, {\it Phase transition and chaos in charged syk model},  {\em
  JHEP} {\bf 07} (2020) 055, [\href{http://arxiv.org/abs/1912.04326}{{\tt
  arXiv:1912.04326}}].

\bibitem{Maldacena:2016hyu}
J.~Maldacena and D.~Stanford, {\it {Remarks on the Sachdev-Ye-Kitaev model}},
  {\em Phys. Rev.} {\bf D94} (2016), no.~10 106002,
  [\href{http://arxiv.org/abs/1604.07818}{{\tt arXiv:1604.07818}}].

\end{thebibliography}\endgroup
\bibliographystyle{JHEP}

\end{document}